\documentclass[preprint,12pt]{elsarticle}

\usepackage{amssymb}

\usepackage{mathptmx}
\usepackage{bm}
\usepackage{amsmath}

\newcommand{\dif}{\mathrm{d}}
\DeclareMathAlphabet{\mathsfsl}{OT1}{cmss}{m}{sl}
\newcommand{\tensor}[1]{\mathsfsl{#1}}

\newtheorem{proposition}{Proposition}[section]

\journal{Communications in Nonlinear Science and Numerical Simulation}

\begin{document}

\begin{frontmatter}
\title{Application of Explicit Symplectic Algorithms to Integration of Damping Oscillators}
\author[author1]{Tianshu Luo}
\author[author2]{Yimu Guo}
\address[author1]{Institute of Applied Mechanics, Department of Mechanics, Zhejiang University, Hangzhou, Zhejiang, 310027,  P.R.China}
\address[author2]{Institute of Applied Mechanics, Department of Mechanics, Zhejiang University, Hangzhou, Zhejiang, 310027,  P.R.China}
\ead[author1]{ltsmechanic@zju.edu.cn}
\ead[author2]{guoyimu@zju.edu.cn}
\date{Received: date / Accepted: date}

\maketitle

\begin{abstract}
In this paper an approach is outlined. With this approach some explicit algorithms can be applied to solve the initial value problem of 
$n-$dimensional damped oscillators. This approach is based upon following structure: for any non-conservative classical mechanical system 
and arbitrary initial conditions, there exists a conservative system; both systems share one and only one common phase curve; and, 
the value of the Hamiltonian of the conservative system is, up to an additive constant, equal to the total energy of the non-conservative 
system on the aforementioned phase curve, the constant depending on the initial conditions.  A key way applying explicit symplectic algorithms 
to damping oscillators is that by the Newton-Laplace principle the nonconservative force can be reasonably assumed to be equal to a function of a component of generalized coordinates $q_i$ along a 
phase curve, such that the damping force can be represented as a function analogous to an elastic restoring force numerically in advance. 
Two numerical examples are given to demonstrate the good characteristics of the algorithms.
\end{abstract}
\end{frontmatter}
\section{Introduction}

Feng\cite{Feng1985,Feng1989,Feng1990,Feng1991},Marsden\cite{Marsden1998},Neri\cite{FNeri1988} and Yoshida\cite{1990PhLA..150..262Y}had 
developed a series of symplectic algorithms for Hamiltonian systems. These algorithms possess some advantages. But it is difficult
 to apply these algorithms to damping dynamical systems, because it has been stated in most classical textbooks that the Hamiltonian 
formalism focuses on solving conservative problems. Damping phenomena is very important in the modeling of
dynamical systems, and can not be avoided. Our aim is to apply some explicit canonical algorithms to nonlinear damping dynamical
 systems, which is generated generally by FE-method. These canonical algorithms reported in this paper can be readily utilized for 
 computing large-scale nonlinear damped dynamical systems.

Betch\cite{Uhlar2010737}\cite{springerlink:10.1007/s00466-003-0516-2}\cite{Betsch20067020} attempted to apply directly some implicit  
algorithms to damping systems. The implicit symplectic algorithms utilized by Betch\cite{Uhlar2010737} possess a few good characteristic, e.g.  
energy-conservation, momentum-consistence, etc... In terms of energy-conservation, implicit symplectic algorithms might be better than 
explicit symplectic ones. But explicit symplectic schemes might be more suitable for nonlinear problems.

If one needs to apply symplectic algorithms to a dissipative system, one must convert the dissipative system into a Hamiltonian system 
or find some relationship between the dissipative system and a conservative one.

In the literature\cite{2009arXiv0906.3062L}, we have stated a proposition describing a relation among a damping dynamical system 
and conservative ones: 
\begin{proposition}
For any non-conservative classical mechanical system and arbitrary initial condition, there exists a conservative system; both systems sharing 
one and only one common phase curve; and the value of the Hamiltonian of the conservative system is equal to the sum of the total energy 
of the non-conservative system on the aforementioned phase curve and a constant depending on the initial condition.
\label{pro:1}
\end{proposition}
In other words, a dissipative ordinary equation and a conservative equation may possess a common particular solution. 
In the next section, an analytical examples are given to explain this proposition. Readers can find the detailed proof 
of Proposition \ref{pro:1} in the reference\cite{2009arXiv0906.3062L} 

In the Literature \cite{Feng1987} a basic explicit canonical integrator is proposed. Based on this basic scheme, Neri\cite{FNeri1988} 
constructed 4-order explicit canonical integrator, and then Yoshida \cite{1990PhLA..150..262Y} proposed a general method to construct 
higher order explicit symplectic integrator. Utilizing the Proposition \ref{pro:1}, we apply this class of explicit canonical 
integrators to damping dynamical systems. This point will be in detail stated in sec. \ref{sec:3}.

\section{One-dimensional Analytical Example}\label{sec:2}
Consider a special one-dimensional simple mechanical system:
\begin{equation}
 \ddot{x}+c\dot{x}=0,
\label{eq:simp_1d}
\end{equation}
where $c$ is a constant. The exact solution of the equation above is
\begin{equation}
 x=A_1+A_2e^{-ct},
\label{eq:sol_1d}
\end{equation}
where $A_1,A_2$ are constants. Differentiation gives the velocity:
\begin{equation}
 \dot{x}=-cA_2e^{-ct}.
\label{eq:sol_1dv}
\end{equation}
From the initial condition $x_0,\dot{x}_0$, we find $A_1=x_0+\dot{x}_0/c, A_2=-\dot{x}_0/c$. Inverting Eq. (\ref{eq:sol_1d}) yields
\begin{equation}
 t=-\frac{1}{c}\ln\frac{x-A_1}{A_2}
\label{eq:tfunc}
\end{equation}
and by substituting into Eq. (\ref{eq:sol_1dv}), such we have
\begin{equation}
 \dot{x}=-c(x-A_1)
\label{eq:dx}
\end{equation}
The dissipative force $F$ in the dissipative system (\ref{eq:simp_1d}) is
\begin{equation}
 F=c\dot{x}.
\label{eq:F}
\end{equation}
Substituting Eq. (\ref{eq:dx}) into Eq. (\ref{eq:F}), the conservative force $\mathcal{F}$ is expressed as
\begin{equation}
 \mathcal{F}=-c^2(x-A_1);
\label{eq:mF}
\end{equation}
Clearly, the conservative force $\mathcal{F}$ depends on the initial condition of the dissipative system (\ref{eq:simp_1d}), in other words,  
an initial condition determines a conservative force. Consequently, a new conservative system yields
\begin{equation}
 \ddot{x}+\mathcal{F}=0\rightarrow \ddot{x}-c^2(x-A_1)=0.
\label{eq:1d_eq_consys}
\end{equation}
The stiffness coefficient in this equation must be negative. One can readily verify that the particular solution (\ref{eq:sol_1d}) 
of the dissipative system can satisfy the conservative one (\ref{eq:1d_eq_consys}). This point agrees with Proposition (\ref{pro:1}).

The potential of the conservative system (\ref{eq:1d_eq_consys}) is 
\[
 V=\int_0^x \left[ -c^2(x-A_1) \right]\dif x =-\frac{c^2}{2}x^2+c^2A_1x 
\]
Therefore the Hamiltonian is
\[
 \hat{H}=T+V=\frac{1}{2}p^2-\frac{c^2}{2}x^2+c^2A_1x,
\]
where $p=\dot{x}$.

\begin{figure}
\begin{center}
\scalebox{0.4}{\includegraphics{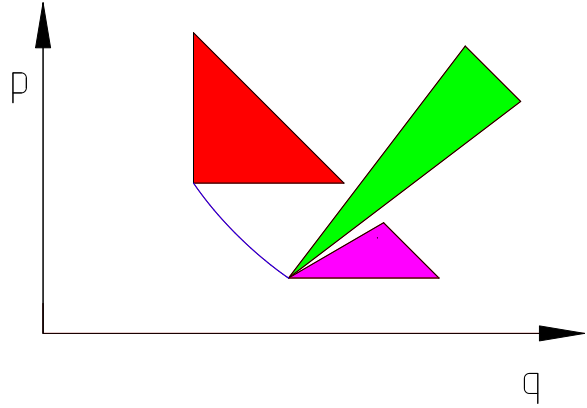}}
\caption{Relationship between nonconservative system (\ref{eq:simp_1d}) and conservative one (\ref{eq:1d_eq_consys})} 
\end{center}
\label{fig:relation}       
\end{figure}

Furthermore, Proposition (\ref{pro:1}) can be depicted by Fig. \ref{fig:relation}. The phase flow of conservative system (\ref{eq:sol_1d}) 
transforms the red area in phase space to the purple area; the phase flow of conservative system (\ref{eq:1d_eq_consys}) transforms 
the red area to the green area. The blue curve in Fig. \ref{fig:relation} illustrates the common phase curve. If one draws more common phase 
curves and phase flows, the picture will like a flower, the phase flow of the nonconservative system likes a pistil and phase flows conservative systems like 
petals.

\section{Modification Symplectic Numerical Schemes}\label{sec:3}
\subsection{Basic Explicit Symplectic Numerical Schemes}
In the paper\cite{Feng1987}\cite{FNeri1988}\cite{1990PhLA..150..262Y} a symplectic algorithm based second kind generation function was stated：
\begin{equation}
 \begin{array}{l}
 {{\bm{p}}^{i + 1}} = {{\bm{p}}^i} - \tau {H_q}({{\bm{p}}^{i + 1}},{{\bm{q}}^i}) \\ 
 {{\bm{q}}^{i + 1}} = {{\bm{q}}^i} + \tau {H_p}({{\bm{p}}^{i + 1}},{{\bm{q}}^i}), \\ 
 \end{array}
\label{eq:base_scheme}
\end{equation}
where the superscript $i$ denotes the $i$-th time node, $\bm{q}$ denotes coordinates and $\bm{p}$ denotes canonical momenta, 
and $H$ denotes Hamiltonian quantity, $H_q=\partial H/\partial \bm{q},\ \ H_p=\partial H/\partial \bm{p}$. 
If the Hamiltonian is seperable, i.e. $H =U(\bm{p})+ V(\bm{q}),V_q=H_q,U_p=H_p$, then the symplectic scheme(\ref{eq:base_scheme}) above becomes 
an explicit symplectic scheme: 
\begin{equation}
\begin{array}{l}
 {{\bm{p}}^{i + 1}} = {{\bm{p}}^i} - \tau {V_q}({{\bm{q}}^i}) \\ 
 {{\bm{q}}^{i + 1}} = {{\bm{q}}^i} + \tau {U_p}({{\bm{p}}^{i + 1}}). \\ 
 \end{array}
\label{eq:basee_exp_scheme}
\end{equation}
For some nonlinear vibration mechanical system, $V_q=\tensor{K}(\bm{q})\bm{q}$.

Let us consider an $n-$dimensional nonlinear oscillator:
\begin{equation}
 \ddot{\bm{q}}+\tensor{C}\dot{\bm{q}}+\tensor{K}\bm{q}=0,
\label{eq:osc}
\end{equation}
where $\tensor{C}$ denotes a non-linear damping coefficient matrix which depends on $\bm{q}$, and $\tensor{K}$ denotes a non-linear stiffness 
matrix which depends on $\bm{q}$ and consists of two parts $\tensor{K}=\tensor{\check{K}}+\tensor{\hat{K}}$($\tensor{\check{K}}$ is a diagonal matrix). 

In accordance with Proposition \ref{pro:1}, a conservative mechanical system was found associated with the dissipative system (\ref{eq:osc}) in 
addition to its initial conditions. Subject to these initial conditions, the dissipative system (\ref{eq:osc}) possesses a common phase curve $\gamma$
 with the conservative system. As in Eq. (\ref{eq:mF}), we can consider that the components of the damping force $\tensor{C}\dot{\bm{q}}$ determine 
the components of a conservative force on the phase curve $\gamma$
\begin{equation}
\begin{array}{ccc}
c_{11}\dot{q}_1=\varrho_{11}(q_1)&\dots&c_{1n}\dot{q}_n=\varrho_{1n}(q_1)\\
\vdots&\ddots&\vdots\\
c_{n1}\dot{q}_1=\varrho_{21}(q_n)&\dots&c_{nn}\dot{q}_n=\varrho_{nn}(q_n).
\end{array}
\label{eq:ex2-4}
\end{equation}
For convenience, this conservative force is assumed to be an elastic restoring force: 
\begin{equation}
\begin{array}{ccc}
\varrho_{11}(q_1)=\kappa_{11}(q_1)q_1&\dots&\varrho_{1n}(q_1)=\kappa_{1n}(q_1)q_1\\
\vdots&\ddots&\vdots\\
\varrho_{n1}(q_1)=\kappa_{n1}(q_n)q_n&\dots &\varrho_{nn}(q_n)=\kappa_{nn}(q_n)q_n .
\end{array}
\label{eq:ex2-5}
\end{equation}

In a similar manner, the components of the non-conservative force $\tensor{\hat{K}}\bm{q}$ are equal to the 
components of a conservative force on the phase curve $\gamma$
\begin{equation}
\begin{array}{ccc}
\hat{K}_{11}q_1=\chi_{11}(q_1)&\dots&\hat{K}_{1n}q_n=\chi_{1n}(q_1)\\
\vdots&\ddots&\vdots\\
\hat{K}_{n1}q_1=\chi_{21}(q_n)&\dots&\hat{K}_{nn}q_n=\chi_{nn}(q_n).
\end{array}
\label{eq:ex2-4a}
\end{equation}
The conservative force can likewise be assumed to an elastic restoring force: 
\begin{equation}
\begin{array}{ccc}
\chi_{11}(q_1)=\lambda_{11}(q_1)q_1&\dots&\chi_{1n}(q_1)=\lambda_{1n}(q_1)q_1\\
\vdots&\ddots&\vdots\\
\chi_{n1}(q_1)=\lambda_{n1}(q_n)q_n&\dots &\chi_{nn}(q_n)=\lambda_{nn}(q_n)q_n .
\end{array}
\label{eq:ex2-5a}
\end{equation}
By an appropriate transformation, an equivalent stiffness matrix $\tensor{\tilde{K}}$ that is diagonal in form can be obtained
\begin{equation}
\tensor{\tilde{K}}_{ii}=\sum_{l=1}^n \kappa_{il}(q_l)+\lambda_{il}(q_l).
\label{eq:ex2-5b}
\end{equation}

Consequently, an $n$-dimensional conservative system is obtained
\begin{equation}
 \bm{\ddot{q}}+(\tensor{\check{K}}+\tensor{\tilde{K}})\bm{q}=0
\label{eq:ex2-6}
\end{equation}
which shares the common phase curve $\gamma$ with the $n$-dimensional damping system described by (\ref{eq:osc}). 
In this paper, the conservative system is called the 'substitute' conservative system.
The Lagrangian of Eqs.(\ref{eq:ex2-6}) is
\begin{equation}
 \hat{L}=\frac{1}{2}\dot{\bm{q}}^T\dot{\bm{q}}-\int_{\bm{0}}^{\bm{q}}(\tensor{\check{K}}\bm{q})^T\dif \bm{q}-
\int_{\bm{0}}^{\bm{q}} (\tilde{\tensor{K}}\bm{q})^T \dif \bm{q},
\label{eq:ex2-7a}
\end{equation}
with the Hamiltonian
\begin{equation}
 \hat{H}=\frac{1}{2}\bm{p}^T\bm{p}+\int_{\bm{0}}^{\bm{q}}(\tensor{\check{K}}\bm{q})^T\dif \bm{q}+
\int_{\bm{0}}^{\bm{q}} (\tilde{\tensor{K}}\bm{q})^T \dif \bm{q},
\label{eq:ex2-7}
\end{equation}
where $\bm{0}$ is the zero vector, and $\bm{p}=\dot{\bm{q}}$. Here $\hat{H}$ in Eq. (\ref{eq:ex2-7}) is the mechanical energy of the 
conservative system (\ref{eq:ex2-6}), because $\int_{\bm{0}}^{\bm{q}} (\tilde{\tensor{K}}\bm{q})^T \dif \bm{q}$
 is a potential function such that $\hat{H}$ is independent of the path taken in phase space.

Subject to a certain initial condition, one need merely to solve the conservative system(\ref{eq:ex2-6}). But one must in advance obtain the 
numerical approximation of the matrix $\tensor{\tilde{K}}$ for a time step, such that one can utilize the algorithm (\ref{eq:basee_exp_scheme}) 
to integrate the conservative system (\ref{eq:ex2-6}) for a time step. One can repeat this process above up to the end. In this way one obtains 
the numerical particular solution of the conservative system (\ref{eq:ex2-6}), which is exactly the numerical particular solution of the damping one.
The he numerical approximation of the matrix $\tensor{\check{K}}$ can be assumed as:
\begin{eqnarray}
\tensor{\tilde{K}}=\left[
\begin{array}{ccc}
 \tilde{K}_{11}&\dots&0\\
\vdots&\ddots&\vdots\\
0&\dots&\tilde{K}_{nn}
\end{array}
\right]  \label{eq:eqv_stiffness}\\
 {\tilde{K}_j}({q_j}^i) = {c_{jl}}\dot q_l^i/{q_j}^i+\hat{K}_{jl}q_l^i/{q_j}^i \nonumber 
\end{eqnarray}
Hence the explicit canonical scheme (\ref{eq:basee_exp_scheme}) can be modified into
\begin{equation}
  \begin{array}{l}
 \tilde{K} _j^i({q_j}^i) = {c_{jl}}\dot q_l^i/{q_j}^i+\hat{K}_{jl}q_l^i/{q_j}^i \\ 
 {p_j}^{i + 1} = {p_j}^i - \tau [{K_j} + \tilde{K} _j^i({q_j}^i)]{q^i}) \\ 
 {q_j}^{i + 1} = {q^i} + \tau {p_j}^{i + 1} \\ 
 \end{array}
\label{eq:1st_exp_schemeeq}
\end{equation}
The scheme above is a one order scheme. Furthermore one can construct higher order explicit canonical schemes utilizing the method 
reported in the literatures\cite{FNeri1988}\cite{1990PhLA..150..262Y}. Now consider a map
from $\bm{z}=\bm{z}(0)$ to $\bm{z}'=\bm{z}(\tau)$:
\begin{equation}
 {\bm{z'}} \approx (\prod\limits_{i = 1}^h {{e^{{r_i}t{\tensor{E}}}}} {e^{{s_i}\tau{\tensor{F}}}}+ O({\tau^{n + 1}})){\tensor{z}},
\label{eq:exponent_form}
\end{equation}
where 
\begin{eqnarray*}
 \bm{z}=\left[\begin{array}{l}\bm{p},\\ \bm{q}\end{array}\right],\bm{z}'=\left[\begin{array}{l}\bm{p}',\\ \bm{q}'\end{array}\right],\\ 
 {\tensor{E}} = \left[ {\begin{array}{*{20}{c}}
   0 & 0  \\
   1 & 0  \\
\end{array}} \right]
{\tensor{F}} = \left[ {\begin{array}{*{20}{c}}
   0 & { - ({\tensor{K}} + {\tensor{\tilde{K} }})}  \\
   0 & 0  \\
\end{array}} \right].
\end{eqnarray*}
In fact Eq.(\ref{eq:exponent_form}) is the succession of the following mappings,
\begin{equation}
 \begin{array}{l}
 {{\tensor{p}}^{j + 1}} = {{\tensor{p}}^j} - {s_i}\tau{V_q}({{\tensor{q}}^j})   \\ 
 {{\tensor{q}}^{j + 1}} = {{\tensor{q}}^j} + {r_i}\tau{U_p}({{\tensor{p}}^{j + 1}}) \\ 
 \end{array}.
\label{eq:h_basee_exp_scheme}
\end{equation}
In reality the difference between the equations above and Eq.(\ref{eq:1st_exp_schemeeq}) is that the coefficients 
$s_i,r_i$ before the time step $\tau$. In the literature \cite{1990PhLA..150..262Y} a generalized method to determine $s_i,r_i$ were 
given. Therefore, the higher order explicit canonical scheme can be represented as:
\begin{equation}
 \begin{array}{l}
 {\tensor{\tilde{K} }}({q^j}) = \left[ {\begin{array}{*{20}{c}}
   {{\tilde{K} _1}(q_1^j)} & {} & 0  \\
   {} &  \ddots  & {}  \\
   0 & {} & {{\tilde{K} _n}(q_n^j)}  \\
\end{array}} \right]       
{\tilde{K} _\alpha }(q_\alpha ^j) = \sum\limits_{l = 1}^n {{c_{\alpha l}}\dot q_l^j/q_\alpha ^j}+\hat{K}_{\alpha l}q_l^i/q_\alpha^i  \\ 
 {\tensor{E}} = \left[ {\begin{array}{*{20}{c}}
   0 & 0  \\
   1 & 0  \\
\end{array}} \right]\;\;\;\;{\tensor{F}} = \left[ {\begin{array}{*{20}{c}}
   0 & { - ({\tensor{K}} + {\tensor{\tilde{K} }})}  \\
   0 & 0  \\
\end{array}} \right] \\ 
 {{\bm{z}}^{j + 1}} = (\prod\limits_{i = 1}^h {{e^{{s_i}\tau{\tensor{F}}}}{e^{{r_i}\tau{\tensor{E}}}}} ){{\bm{z}}^j} \\ 
 \end{array}\
\label{eq:h_exp_scheme}
\end{equation}

\section{Numerical Examples}
Two examples will be given to shown this numerical method\ref{eq:h_exp_scheme}. 
\subsection{The First Example}
\begin{figure}
 \scalebox{0.8}{\includegraphics{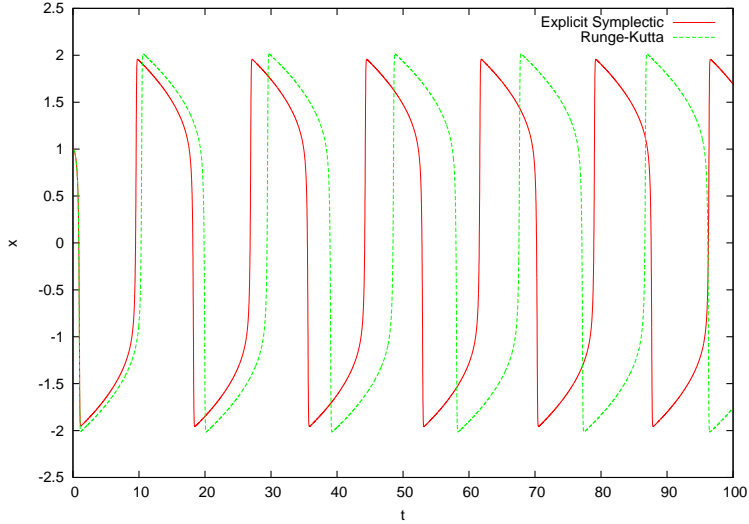}}
\caption{The resonance of the Van Der Pol's oscillator }
\label{fig:x_vdp}       
\end{figure}

\begin{figure}
 \scalebox{0.8}{\includegraphics{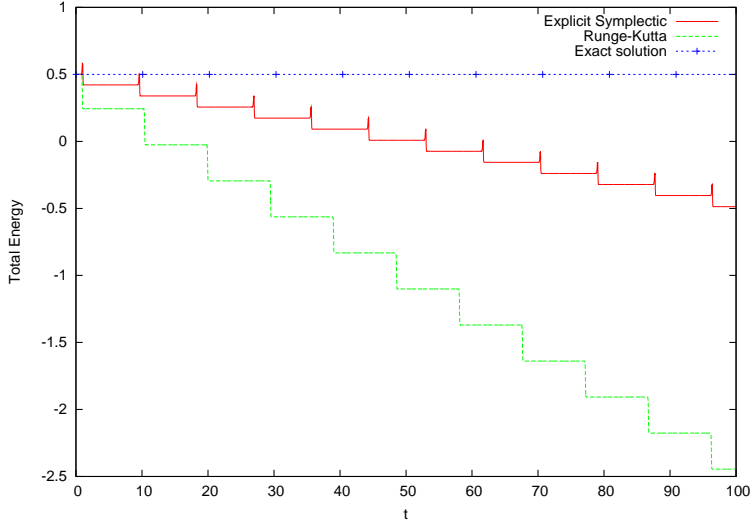}}
\caption{The total energy of the Van Der Pol's oscillator }
\label{fig:energy_vdp}       
\end{figure}

To begin, we consider a Van Der Pol's oscillator
\begin{equation}
\ddot x + \mu \dot x({x^2} - 1) + x = 0,
\label{eq:vdp}
\end{equation}
where $\mu  = 10$. The initial conditions are given by ${x_0} = 1,\;\;\dot x = 0$.
We employee the $4-$order explicit symplectic method (\ref{eq:h_exp_scheme}) with coefficients 
\begin{eqnarray*}
s_1=s_4=[2+(\sqrt[3]{2}+1/\sqrt[3]{2})]/6,\ \ s_2=s_3=[1-(\sqrt[3]{2}+1/\sqrt[3]{2})]/6, \\
\ \ r_1=r_3=[2+(\sqrt[3]{2}+1/\sqrt[3]{2})]/3,\ \ r_2=-[2+(\sqrt[3]{2}+1/\sqrt[3]{2})]/3,\ \ r_4=0,
\end{eqnarray*}
and classical explicit $4-$order Runge-Kutta method to compute the resonance of the Van Der Pol's oscillator (\ref{eq:ex_duffing}) respectively, 
then employ a same method to integrate the results to the total energy, which is the sum of the mechanical energy 
and the work done by damping forces in the system (\ref{eq:vdp}). The both methods are run with the same step size $\tau=0.01$. 
The resonance is shown in Fig. \ref{fig:x_vdp}, and the total energy is shown 
 in Fig. \ref{fig:energy_vdp}. 

It is aparent from Fig. \ref{fig:energy_vdp} that the explicit symplectic method (\ref{eq:h_exp_scheme}) has qualitatively different behavior to the 
Runge-Kutta method. The energy divergence between the explicit symplectic method and the exact solution is smaller than that between Runge-Kutta method and 
the exact solution. The energy divergence between the explicit symplectic method and Runge-Kutta method increases with the time evolution. Due to the 
increasement of the energy, the phase difference between both the results in Fig. \ref{fig:x_vdp} increases also with the time evolution.

\subsection{The Second Example}
In the second example, we consider a $2-$dimensional damped nonlinear Duffing oscillator
\begin{equation}
\begin{array}{l}
 2\ddot{q}_1+0.1\dot{q}_1+(2+0.1q_1^2)q_1+q_2=0\\
 3\ddot{q}_2+0.2\dot{q}_2+q_1+(2+0.2q_2^2)q_2=0,
\end{array}
\label{eq:ex_duffing}
\end{equation}
with the initial conditions $q_1=0,\ \ q_2=0,\ \ \dot{q}_1=0,\ \ \dot{q}_2=1$. The program of the both methods with the step size $\tau=0.01$ are 
carried out to simulate Eq. (\ref{eq:ex_duffing}). The resonance is shown in Fig. \ref{fig:x_vdp}, the numerical solution of the total energy 
is shown in Fig.\ref{fig:te}.

There is only tiny difference between resonance results of the two methods, correspondingly, the difference among the total energy 
obtained by the numerical methods and anlytical methods is very tiny. As numerical examples in the other literatures\cite{Kane2000}, that 
explicit Runge-Kutta method must cause numerical pseudo dissipation which might be positive or negative. The difference between our 
numerical examples and the examples in the literature\cite{Kane2000} is the total energy in our examples and the mechanical energy in 
their examples\footnote{Fig.6.1 in the literature\cite{Kane2000}}.
 \begin{figure}
 \scalebox{0.8}{\includegraphics{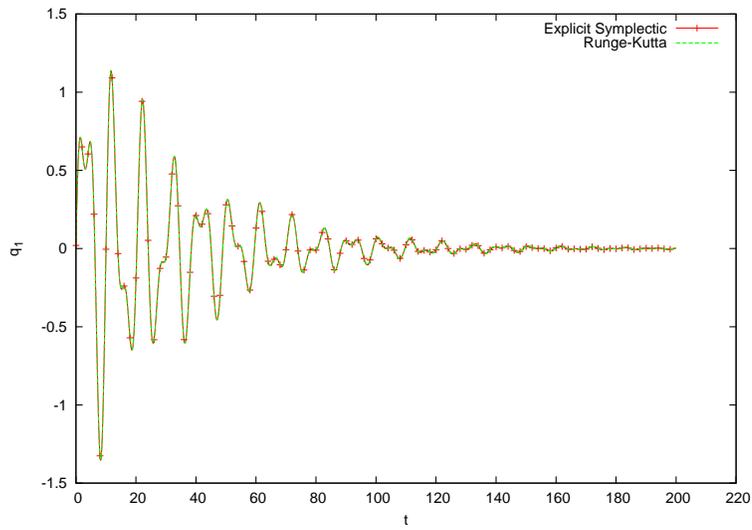}}
\caption{The $1$-th displacement of the damped Duffing oscillator }
\label{fig:q1g}       
\end{figure}

 \begin{figure}
 \scalebox{0.8}{ \includegraphics{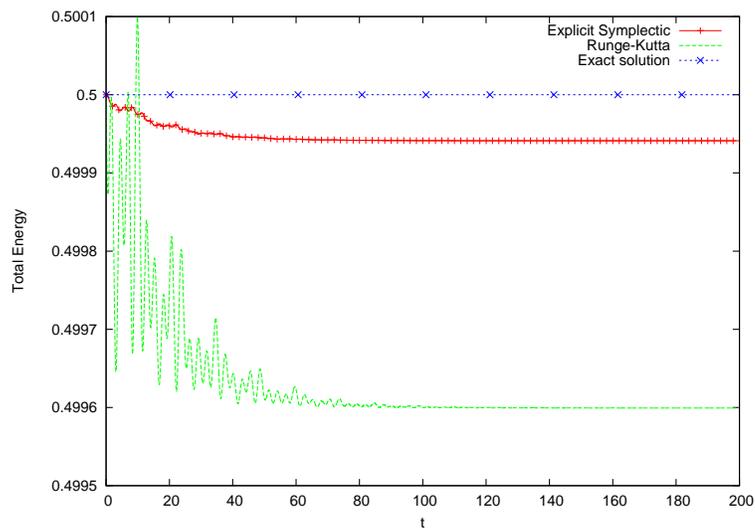}}
\caption{Total energy of the damped dissipative oscillator }
\label{fig:te}       
\end{figure}

\section{Conclusions}
We have introduced  a class of explicit symplectic algorithms to dissipative mechanical systems successfully, by changing 
these algorithms into the scheme.(\ref{eq:h_exp_scheme}). Because the algorithms (\ref{eq:h_exp_scheme}) are explicit and 
possess good energy preserving characteristics, the explicit symplectic algorithms (\ref{eq:h_exp_scheme}) is quite suitable 
for long term integration of arbitrary dimensional nonlinear dissipative mechanical systems.

\end{document}